\documentclass[
twocolumn,
amsfonts,
showpacs,
superscriptaddress,
amsmath,
amssymb,
aps,
longbibliography,
pnas,
noeprint
]{revtex4-2}

\usepackage{graphicx}% Include figure files
\usepackage{bm}% bold math
\usepackage[dvipsnames]{xcolor}
\usepackage[english]{babel}
\usepackage{dsfont}
\usepackage{comment}
\usepackage[caption=false]{subfig}
\usepackage[colorlinks,%
            linkcolor=BrickRed,%
            citecolor=MidnightBlue,%
            urlcolor=MidnightBlue]{hyperref}

\newcommand{\bc}{\begin{center}}
\newcommand{\ec}{\end{center}}
\def\ba#1{\begin{array}{#1}\displaystyle}
\newcommand{\ea}{\end{array}}

\newcommand{\beq}{\begin{equation}}
\newcommand{\eeq}{\end{equation}}
\newcommand{\beqa}{\begin{eqnarray}}
\newcommand{\eeqa}{\end{eqnarray}}

\newcommand{\n}{\nonumber\\}
\newcommand{\bi}{\begin{itemize}}
\newcommand{\ei}{\end{itemize}}
\newcommand{\p}{\partial}

\newcommand{\ii}{{\rm i}}

\newcommand{\dd}{{\rm d}}

\begin{document}

\title{Anomalous current fluctuations from Euler hydrodynamics}

\author{Takato Yoshimura}
\email{takato.yoshimura@physics.ox.ac.uk}
\affiliation{All Souls College, Oxford OX1 4AL, U.K.}
\affiliation{Rudolf Peierls Centre for Theoretical Physics, University of Oxford, 1 Keble Road, Oxford OX1 3NP, U.K.}

\author{\v{Z}iga Krajnik}
\email{ziga.krajnik@nyu.edu}
\affiliation{Department of Physics, New York University, 726 Broadway, New York, NY 10003, USA}
\begin{abstract}
We consider the hydrodynamic origin of anomalous current fluctuations in a family of stochastic charged cellular automata. Using ballistic macroscopic fluctuation theory, we study both large and typical fluctuations of the charge current and reproduce microscopic results available for the deterministic single-file limit of the models. Our results indicate that in general initial fluctuations propagated by Euler equations fully characterize anomalous fluctuations on both scales. In the stochastic case, we find an additional contribution to typical fluctuations and conjecture the functional form of the typical probability distribution, which agrees with numerical simulations.
\end{abstract}

\maketitle

\section{Introduction}
Statistical physics aims to describe universal behaviors emerging from interactions of a large number of microscopic degrees of freedom. While traditionally the main focus has been equilibrium physics, in recent years it has been appreciated that out-of-equilibrium properties of the system can also display rich universal behaviors. A prominent tool that describes such physics is macroscopic fluctuations theory (MFT) \cite{MFT}, which captures large fluctuations of driven diffusive systems. The validity of MFT has been verified for a variety of systems \cite{Derrida2005,Derrida2009,Krapivsky2015,Mallick2022} and also on quantum simulators \cite{wienand2023emergencefluctuatinghydrodynamicschaotic}, but large fluctuations in these systems usually behave in a regular way with finite scaled cumulants of observables. This naturally raises the question of whether there exist systems that break such regularity and support new dynamical phenomena.

Recent numerical studies have found dynamical criticality and anomalous spin fluctuations in integrable spin chains \cite{Krajnik2022_1,Krajnik2024_1}, which have also been probed experimentally using superconducting qubits \cite{Rosenberg_2024}. While the mechanism responsible for anomalous spin fluctuations remains to be clarified, an exact calculation of fluctuations in a deterministic charged cellular automaton \cite{Krajnik2022} provided a basic example of dynamical criticality in a deterministic many-body system. The procedure has been generalized to a wider class of single-file dynamics, which was explored in Refs. \cite{Krajnik_Schmidt_Pasquier_Prosen_Ilievski_2024,krajnik2024singlefile,Altshuler_Konik_Tsvelik_2006,Feldmeier2022,Kormos2022,Bidzhiev2022}. Furthermore, anomalous fluctuations have also been observed in the easy-axis regime of an integrable spin chain \cite{Krajnik2024_1,SarangKhemani2024} and other systems including Dirac fluids \cite{gopalakrishnan2024nongaussian} and stochastic processes \cite{mcculloch2024ballisticmodessourceanomalous}, hinting at a more general setting supporting anomalous fluctuations. Anomalous spin transport in the isotropic Heisenberg chain has recently been under intense theoretical and experimental scrutiny \cite{Znidaric2011,Ljubotina2017,Ljubotina2019a,Das2019a,Krajnik2020,Krajnik2020a,Weiner2019,Dupont2020,Jepsen2020,Scheie2021,Wei2022}. While the statistics of magnetization transfer is fully captured by the Baik-Rains subuniversality at the level of two-point quantities \cite{takeuchi2024}, higher-point quantities are distinct from the Kardar-Parisi-Zhang (KPZ) univerality class \cite{Kardar1986,Corwin2012,Quastel2015,Takeuchi2018} by symmetry \cite{Krajnik2022_1}, leading to {\it partial} KPZ physics.

Motivated by these, in this Letter, we provide a full hydrodynamic framework to understand the emergence of anomalous charge fluctuations in a family of stochastic charged cellular automata (SCCA), which was first introduced in Ref.~\cite{Klobas_Medenjak_Prosen_2018} and reduces to the model studied in Ref.~\cite{Krajnik2022} in the single-file (deterministic) limit. The main tool for the analysis is ballistic MFT (BMFT) \cite{BMFT,PhysRevLett.131.027101}, which is a generalization of MFT for systems that support ballistic transport, and we employ it to characterize both typical and large fluctuations.
The simple three-mode hydrodynamic structure allows for an exact solution of BMFT equations, showing that large charge fluctuations are identical in the whole family of SCCA. While BMFT is in principle applicable only to large fluctuations, we also demonstrate that an approach based on the same idea, which is that current fluctuations are solely determined by initial fluctuations transported by Euler hydrodynamics, accurately describes typical fluctuations in the single-file limit, recovering the microscopic result \cite{Krajnik2022}.
%Building upon the idea that charge fluctuations in the single-file limit stem solely from hydrodynamically transported fluctuations of the initial state, we recover charge distributions on the typical scale obtained in Ref.~\cite{Krajnik2022}, demonstrating that Euler scale physics is sufficient to describe sub-ballistic phenomena. 
Interestingly, unlike large fluctuations, it turns out that typical fluctuations depend on the stochasticity of dynamics. Based on numerical simulations, we conjecture a form of the corresponding typical distribution at half-filling.

\begin{figure}[t!]
    \centering
        \includegraphics[width=\columnwidth]{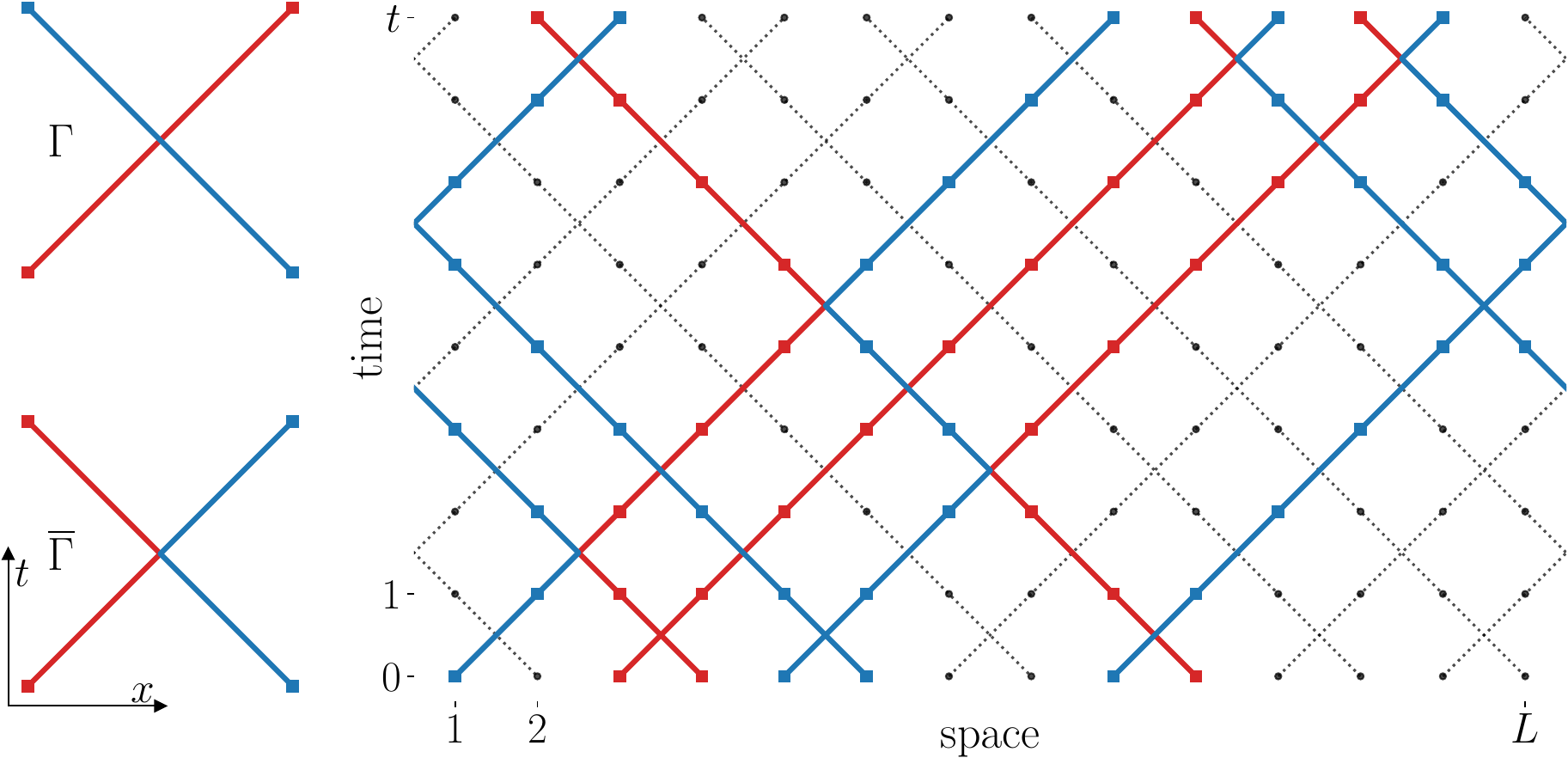}
        \caption{(left) Stochastic particle scattering in the two-particle sector of the local two-body map $\Phi$. Particles with positive/negative charge (red/blue squares) either cross (top left) with probability $\Gamma$ or are elastically reflected (bottom left) with probability $\overline \Gamma$. Particle worldlines (colored lines) shown for clarity. (right) Many-body dynamics of charged particles and vacancies (black circles) in discrete space-time.  Particles move along diagonals except when they encounter another particle when they scatter stochastically.}
	\label{fig:dynamics}
\end{figure}

\section{The model}
We consider a family of SCCA on a one-dimensional lattice of $L \in 2\mathbb{N}$ sites. The lattice configuration at fixed time $t \in \mathbb{Z}$ is given by strings ${\bf s}^t \equiv s^t_{L}s^t_{L-1} \ldots s^t_1$
of symbols $s^t_x \in \{\emptyset, -, +\}$, which correspond to vacancies and negatively or positively charged particles, respectively.
Local symbol dynamics are given by a one-parameter stochastic two-body map $(s_L', s_R') = \Phi(s_L, s_R)$ which encodes the following dynamical rules $(\emptyset, \emptyset) \rightarrow (\emptyset, \emptyset)$, $(\emptyset, c) \leftrightarrow (c, \emptyset)$ while $(c, c') \to (c', c)$ with probability $ 0 \leq \Gamma \leq 1$ and $(c, c') \to (c, c')$ with probability $\overline \Gamma \equiv 1-\Gamma$ where $c,c' \in \{-,+\}$, see Figure~\ref{fig:dynamics}. 
% \zk{We note that the dynamics can be understood as a composition of a stochastic six-vertex model in the particle-particle sector with free dynamics in the remaining sectors. - CAN BE CUT}
%, see the Supplemental Material (SM) \cite{SM} for details.
The many-body dynamics is realized as a `brickwork' circuit obtained by imposing periodic boundary conditions, $s^t_{x} = s^t_{L+x}$, and coupling alternating sites as ${\bf s}^{2t+2} = \Phi^{\rm odd}({\bf s}^{2t+1}) $ and  $ {\bf s}^{2t+1} = \Phi^{\rm even}({\bf s}^{2t})$, where \begin{equation}
\Phi^{\rm odd}=\prod_{x=1}^{L/2}\Phi_{2x-1,2x}, \qquad  \Phi^{\rm even}=\prod_{x=1}^{L/2}\Phi_{2x,2x+1}
\end{equation}
and $\Phi_{x,x+1}$ acts non-trivially only on sites $x$ and $x+1$, see Figure~\ref{fig:dynamics} for a sample many-body trajectory. For $\Gamma=0$ and $\Gamma=1$ the dynamics is deterministic and corresponds to single-file \cite{Medenjak_Klobas_Prosen_2017,Medenjak2019,Krajnik2022,Krajnik_Schmidt_Pasquier_Prosen_Ilievski_2024} and free dynamics respectively.

While the SCCA are super-integrable and support an exponential (in $L$) number of local conserved quantities \cite{Gombor2022}, we presently restrict our study to a (closed) subset of conserved charges, comprising the number of right and left moving particles $\hat{Q}_\pm$ and the total charge $\hat{Q}_c$, all of which are given as sums of local densities $\hat{Q}_i({\bf s}) = \sum_{x=1}^{L} \hat{q}^{i}_{x, t}({\bf s})$ where $i \in \mathcal{I} \equiv \{+, -, c\}$ and $\hat \bullet$ indicates a microscopic quantity. Explicitly, the local charge densities read  $\hat  q^i_{x,t}(\emptyset) = 0$, $\hat q^c_{x,t}(\pm) = \pm 1$, $\hat  q^-_{x,t}(\pm) = \delta_{x+t, {2\mathbb{Z}+1}}$ and $\hat  q^-_{x,t}(\pm) = \delta_{x+t, {2\mathbb{Z}}}$ 
with corresponding local current densities $\hat \jmath^i_{x, t+1} = (-1)^x (\hat q^i_{x, t+1+(-1)^x}  -  \hat q^i_{x, t+1})/2$
that satisfy discrete continuity equations of the form
\begin{equation}
	\frac{1}{2}\left(\hat q^i_{x, t+2} - \hat q^i_{x, t}\right) + \hat \jmath^i_{x+1, t+1} - \hat \jmath^i_{x, t+1} = 0.
    \label{local_continuity}
\end{equation}
%for details on charge densities and discrete space-time continuity relations.
We consider generalized Gibbs ensembles (GGEs) of the form $\mathbb{P}({\bf s}) = \exp \left[\beta^i \hat{Q}_i({\bf s})\right]/\sum_{\bf s'} \exp \left[\beta^i \hat{Q}_i({\bf s'}) \right],$ where repeated upper and lower indices indicate summation.
The average of an observable $o({\bf s})$ is accordingly $\langle o \rangle = \sum_{{\bf s}} \mathbb{P}({\bf s})\, o({\bf s}).$ Identifying the ensemble parameters as 
$\rho_\pm = [1 + e^{-\beta_\pm}/(2 \cosh \beta_c)]^{-1}, b =  \tanh \beta_c$,
the ensemble factorizes
$\mathbb{P}({\bf s}) = \prod_{x=1}^{L/2}p_-(s_{2x+1}) p_+(s_{2x})$
in terms of normalized one-site measures $p_-(\pm) = \rho_- \frac{1\pm b}{2}$, $p_+(\pm) = \rho_+ \frac{1\pm b}{2}$, $p_\pm(\emptyset) = \overline \rho_\pm$ with $0 \leq \rho_\pm \leq 1$ and $\overline \rho_\pm \equiv 1 - \rho_\pm$ the densities of right/left movers and vacancies on right/left running diagonals, respectively.

\section{Hydrodynamics}
The hydrodynamics of the system turns out to have a simple structure with densities of left and right moving particles  $\varrho_\pm/2$ and charge $\varrho_c$ forming a {\it closed set of hydrodynamic equations} (here space-time dependent densities $\varrho_i$ are distinguished from the static ones $\rho_i$). To study charge transport (in the chosen GGE) it is then sufficient to consider the hydrodynamics of only these three charges unlike in standard integrable systems where all charges must be included. 
%Note that the situation is similar to the case of hydrodynamics of another cellular automaton called Rule 54 (or equivalently that of $T\Bar{T}$-deformed conformal field theories) \cite{}, where left and right moving solitons constitute a closed set of hydrodynamic equations.

The three-mode hydrodynamic equations of the SCCA are obtained by evaluating the flux Jacobian. To this end we first compute the static susceptibility matrix 
$
\mathsf{C}_{ij} = L^{-1} \sum_{x,x'=1}^L \langle \hat  q^i_{x,t} \hat  q^{j}_{x',t} \rangle^c
$, with $\langle xy\rangle^c \equiv \langle xy\rangle - \langle x \rangle \langle y \rangle$ as
\begin{equation}
	\mathsf{C}_{ij} = \frac{1}{2}
	\begin{bmatrix}
	\rho_+ \overline \rho_+ & 0 & b\rho_+ \overline \rho_+\\
	0 & \rho_- \overline \rho_- & b\rho_- \overline \rho_-\\
	b\rho_+ \overline \rho_+ & b\rho_- \overline \rho_- & 2 \rho - 2b^2(\rho^2 + p^2)
\end{bmatrix},
\end{equation}
where $\rho \equiv (\rho_+ + \rho_-)/2$ and $p \equiv (\rho_+ - \rho_-)/2$ are the average particle and momentum density per lattice site.
%We henceforth work in the thermodynamic limit by by sending $L \to \infty$.
To compute the the static charge-current correlator $ 
	\mathsf{B}_{ij} = L^{-1} \sum_{x,x'=1}^L \langle \hat q^i_{x,t}\, \hat \jmath^{j}_{x',t+1} \rangle^c$
we observe that left/right movers move freely along diagonals of distinct sublattices with velocities $\pm 1$ which together with the symmetry $\mathsf{B}_{ij} = \mathsf{B}_{ji}$ immediately yields the relations $
	\mathsf{B}_{\pm \mp} = 0,\ \mathsf{B}_{\pm \pm} = \pm \rho_\pm \overline \rho_\pm/2, \
    \mathsf{B}_{\pm c} = \pm \rho_\pm \overline \rho_\pm b/2$.
To evaluate the $\mathsf{B}_{cc}$ element, we note that the local dynamics generate a non-zero charge current only in the following six configurations
$(\emptyset, \pm) \leftrightarrow (\pm, \emptyset)$, $(+, -) \leftrightarrow (-, +)$.
However, the latter two configurations enter into the average with identical ensemble weights while their current values are oppositely signed so that their contributions identically cancel. This also shows that $\mathsf{B}$ is independent of the crossing parameter $\Gamma$. Evaluating the remaining non-zero charge current configurations, we find
$\mathsf{B}_{cc}
    %= \frac{1}{2}(\rho_+ - \rho_-)[1-b^2(\rho_+ + \rho_-)] 
	= p(1-2b^2 \rho)$,
which gives the full static charge-current correlator
\begin{equation}
	\mathsf{B}_{ij} = 
    \frac{1}{2}
	\begin{bmatrix}
		\rho_+ \overline \rho_+ & 0 & \rho_+ \overline \rho_+ b\\
		0 & -\rho_- \overline \rho_-  & -\rho_- \overline \rho_- b \\
		\rho_+ \overline \rho_+ b  &  -\rho_- \overline \rho_- b  & 2p(1-2b^2 \rho)
	\end{bmatrix}.
\end{equation}
With the matrices $\mathsf{B}$ and $\mathsf{C}$ at our disposal, we can obtain the flux Jacobian \eqref{eq:flux_jacobian} by inverting the relation $\mathsf{B}=\mathsf{A}\mathsf{C}$, yielding
\begin{equation}\label{eq:flux_jacobian}
    [\mathsf{A}[\underline{\rho}]]_i^{~j}=\frac{\p j_i[\underline{\rho}]}{\p\rho_j}=\begin{pmatrix}
    1 & 0&0\\
    0&-1&0\\
    b\rho_-/\rho&-b\rho_+/\rho& v
    \end{pmatrix}
\end{equation}
 where three modes are labeled by $\rho_1=\rho_+/2$, $\rho_2=\rho_-/2$, $\rho_3=\rho_c$ and the charge velocity $v$ is given by $v=p/\rho$
 %with $\rho=\rho_1 + \rho_2$, $p=\rho_1 - \rho_2$ the particle density and momentum respectively
 and $j_i[\underline{\rho}]$ is the average current defined via the continuity equation $\p_t\varrho_i+\p_xj_i[\underline{\varrho}]=0$. Here, $\varrho_i$ is a space-time dependent density and the underline implies that the quantity is multicomponent. Charge inertness and chiral factorization of dynamics amount to free ballistic propagation of the left and right movers $\p_t(\varrho_\pm/2)\pm\p_x(\varrho_\pm/2)=0$. On the other hand, charge transport is affected by particle dynamics and the Euler equation is of the form
\begin{equation}\label{eq:charge_hydro}
    \p_t\varrho_c+\p_x j_c=0,\quad j_c=v\varrho_c.
\end{equation}
% Importantly, the velocity $v$ does not depend on the charge field $\varrho_c$ itself, which is a crucial requirement for the emergence of anomalous fluctuations.
%The diffusion constants $D_c^{~i}$ in the single-file limit $\Gamma=0$  are $D_c^{~c}=(1-v^2)\overline \rho/\rho$  and $D_c^{~\pm}=-bD_c^{~c}$. \zk{Comment on $\Gamma$ dependence.}

%A few comments on the structure of Eq.~\eqref{eq:charge_hydro} are in order.
A system of three-mode hydrodynamic equations similar to Eq.~\eqref{eq:charge_hydro} along with the free chiral equations for $\varrho_\pm$ was also recently derived for Dirac fluids by keeping only the most relevant terms based on symmetry considerations in Ref.~\cite{gopalakrishnan2024nongaussian}. In the present case, the hydrodynamic equations are obtained using the exact flux Jacobian Eq.~\eqref{eq:flux_jacobian}, and interestingly, are completely insensitive to the crossing probability $\Gamma$. This being said, typical fluctuations appear to depend nontrivially on $\Gamma$. Before discussing this dependence, we first study large fluctuations of the SCCA, which can be fully characterized using only Euler hydrodynamic data.

\section{Anomalous charge FCS from BMFT}
The central object in the study of fluctuations is the generating function $\langle e^{\ii \lambda \hat{J}(t)}\rangle=\int\dd\lambda\,e^{\ii\lambda J}\mathcal{P}(J|t)$ given by the probability distribution $\mathcal{P}(J|t)$ of time-integrated charge current $\hat{J}(t)=\int_0^t\dd t'\,\hat{\jmath}_{c}(0,t')$. 
%In equilibrium odd moments of the time-integrated current vanish due to time-reversal symmetry $\langle[\hat{J}(t)]^{2n-1}\rangle=0$, 
While the scale on which {\it typical fluctuations} take place $\hat{J}(t)\sim t^{1/{2z}}$ is set by the variance $\langle[\hat{J}(t)]^2\rangle^c\sim t^{1/z}$, large (atypical) fluctuations occur on the scale $\hat{J}(t)\sim t$ with a probability distribution governed by a large deviation principle $\mathcal{P}({J}=jt|t)\asymp e^{-tI(j)}$  where $I(j)$ is the large deviation rate function. The rate function is related to the {\it scaled cumulant generating function} (SCGF) $F(\lambda)=\lim_{t\to\infty}t^{-1}\log\langle e^{\lambda \hat{J}(t)}\rangle$ via the Legendre transform $I(j)=\max_{\lambda \in \mathbb{R}}[\lambda j-F(\lambda)]$. We now compute the charge SCGF in the SCCA explicitly using BMFT. 

BMFT is a hydrodynamic large deviation theory, which was introduced in Ref.~\cite{BMFT} by generalizing the idea of MFT to systems that support ballistic transport, e.g. integrable systems. Following the principle of MFT, BMFT posits that large dynamical fluctuations at the Euler scale are entirely fixed by Euler hydrodynamics, namely fluctuating densities of the conserved charges transported by the Euler equation. However, the crucial difference from the standard MFT is that there is no source of fluctuations other than those in the initial condition at the Euler scale. This indicates that large dynamical fluctuations at later times can be obtained simply by propagating the initial fluctuations according to the Euler equation.

To implement this idea, we introduce the {\it fluctuating} mesoscopic densities evaluated at the Euler-scale coordinates $(\tau x,\tau t)$, i.e. $\varrho_i(x,t)=\hat{q}^i_{\tau x,\tau t}$ for $\tau\gg1$. To determine how the fields $\varrho_i(x,t)$ fluctuate at later times, we need to characterize the initial fluctuations carried by $\varrho_i(x,0)$; the probability distribution of the initial fields is known to be given by $\mathbb{P}[\underline{\varrho}(\cdot,0)]\asymp e^{-\tau\mathcal{F}[\underline{\varrho}(\cdot,0)]}$, where $\mathcal{F}[\underline{\varrho}(\cdot,0)]$ is the density large deviation function (see Eq.~\eqref{eq:bmft_initial_fluc} in Appendix.~\ref{sec:deriv_bmft}). BMFT then asserts that the moment generating function $\langle e^{\lambda \hat{J}(\tau)}\rangle$ can be expressed as 
\begin{equation}\label{eq:SCGF_path2}
  \langle e^{\lambda \hat{J}(\tau)}\rangle=\int_{(\mathbb{S})}\mathcal{D}\underline{\varrho}(\cdot,\cdot)\mathcal{D}\underline{H}(\cdot,\cdot)e^{-\tau S[\underline{\varrho},\underline{H}]},
\end{equation}
where the functional integral is performed over the space of functions supported on the space-time region $\mathbb{S}=\mathbb{R}\times[0,1]$. The auxiliary field $\underline{H}$ in Eq.~\eqref{eq:SCGF_action} ensures that the fluctuating fields always satisfy the hydrodynamic equations, which in turn gives the BMFT action $S[\underline{\varrho},\underline{H}]$
\begin{equation}\label{eq:SCGF_action}
    S[\underline{\varrho},\underline{H}]=-J(1)+\mathcal{F}[\underline{\varrho}(\cdot,0)]+\int_{\mathbb{S}}\dd x\dd t\,H^i(\p_t\varrho_i+\p_xj_i),
\end{equation}
where $J(1)=\int_0^1\dd t\,j_c(0,t)$ is the time-integrated current. The large time asymptotics of $\langle e^{\lambda \hat{J}(\tau)}\rangle$ is thus given by the saddle point of Eq.~\eqref{eq:SCGF_path2}, and the solution to the \emph{BMFT equations}  that characterizes the saddle point, allows us to evaluate the SCGF exactly. In the SCCA it is possible to carry out the analysis explicitly (see Appendix.~\ref{sec:deriv_bmft} for the derivation),  resulting in the SCGF
\begin{equation}\label{eq:SCGF_exact}
    F(\lambda)=\tfrac{1}{2} \log \left[ 1 + \Delta^2(\mu_b + \mu_b^{-1} -2)  \right]
\end{equation}
where $\Delta^2=\rho(1-\rho)$ and $\mu_b = \cosh \lambda + |b|\sinh |\lambda|$. This coincides with the microscopic result obtained in \cite{Krajnik2022} up to prefactor $1/2$ due to a different convention for the unit of time.

As also observed in Ref.~\cite{Krajnik2022}, the generating function $\langle e^{\ii\lambda \hat J(t)}\rangle$ %Eq.~\eqref{eq:SCGF_exact}
is anomalous with singular higher scaled cumulants $c_{2n>4}=\lim_{t\to\infty}t^{-1}\langle [\hat{J}(t)]^{2n}\rangle^c$. BMFT clarifies the physical origin of this anomalous behavior: since the normal mode $b(x,t)$ of the charge $\varrho_c(x,t)$ travels along the ray $x=0$, fluctuations along $x=0$ proliferate and render clustering of connected charge current correlations algebraic in time \cite{Doyon2019}. While such a scenario generally results in superdiffusion \cite{Spohn2014}, it is avoided here due to the charge conjugation symmetry of the system \cite{Krajnik_Schmidt_Pasquier_Prosen_Ilievski_2024}.  In general, we expect that, for a charge in the system to show anomalous fluctuations, it has to have a vanishing velocity in equilibrium (usually afforded by a $\mathbb{Z}_2$ symmetry of the charge) while other modes should be independent of their own velocities (i.e. no self-coupling), which would otherwise induce superdiffusion of the charge as in the heat mode with the $5/3$-Levy exponent in anharmonic chains \cite{Spohn2014}. 
%It is also reasonable to assume that anomalous fluctuations are robust away from charge inertness, but a more precise analysis of such a situation is beyond the scope of the present Letter.

Note that Eq.~\eqref{eq:SCGF_exact} is valid for any value of $b\in(-1,1)$, including $b=0$, where the dynamical exponent becomes $z=2$. On the other hand, for typical fluctuations the two cases $b\neq0$ and $b=0$ have to be treated separately since the dynamical exponents in the two cases differ. However, it turns out that, in the single-file (i.e., deterministic) limit $\Gamma=0$, the underlying mechanism, namely that initial fluctuations evolving according to the Euler equations fully determine charge current fluctuations, remains true even for typical fluctuations.

\begin{figure*}[t!]
	\centering
	\includegraphics[width=\linewidth]{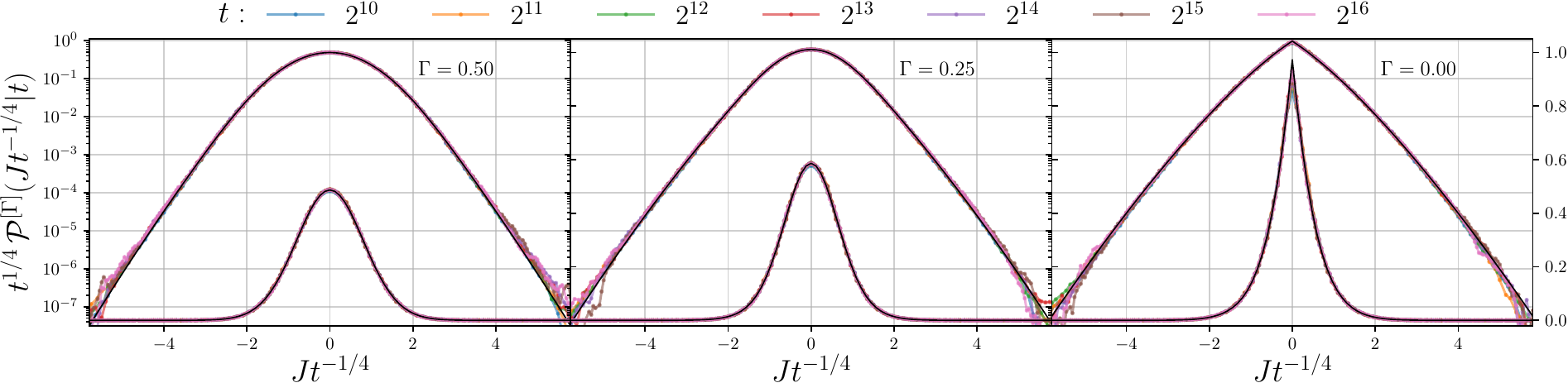}
	\caption{Finite-time typical distributions of the integrated charge current $t^{1/4} \mathcal{P}^{[\Gamma]}(Jt^{-1/4}|t)$ (colored lines) in linear and logarithmic scales at different $\Gamma$ compared against distributions $\mathcal{P}^{[\Gamma]}_{\rm typ}$ \eqref{generalized_dist} (black lines) with $\omega$ and $\sigma$ as fitting parameters. %best-fit values reported in the SM \cite{SM}. 
    Simulation parameters: $\rho=1/2$, $b=0$, $L = 2^{20}$, $t_{\rm max}=2^{16}$, $5\times 10^3$ samples.}
	\label{fig:typical_distributions}
\end{figure*}

\section{Typical fluctuations of deterministic dynamics from Euler hydrodynamics}
The probability distribution of typical charge fluctuations $\hat J(t)\sim t^{1/2z}$ is given by
\begin{equation}\label{eq:prob_typical}
    \mathcal{P}(Jt^{-1/2z}|t)= \int_\mathbb{R}\frac{\dd\lambda}{2\pi}\,e^{-\ii\lambda J t^{-1/2z}}\langle e^{\ii\lambda \hat J(t)}\rangle,
\end{equation}
which converges to $\mathcal{P}_\mathrm{typ}(j)=\lim_{t\to\infty}t^{1/2z}\mathcal{P}(J=jt^{1/2z}|t)$. We now evaluate Eq.~\eqref{eq:prob_typical} using only Euler hydrodynamic data, focusing on the single-file limit $\Gamma=0$, which has been discussed extensively in Refs.~\cite{Krajnik2022,Krajnik_Schmidt_Pasquier_Prosen_Ilievski_2024,krajnik2024singlefile}.

 Recall that, the main ideas of both MFT and BMFT, which describe large fluctuations, hinge on the postulate that the generating function $\langle e^{\ii\lambda \hat J(t)}\rangle$ can be expressed solely in terms of diffusive- or Euler-scale variables $\varrho_i(x,t)$, respectively, whose fluctuations are of order $O(1)$.  Importantly, these fluctuating variables originate from atypical initial fluctuations described by
 Eq.~\eqref{eq:bmft_initial_fluc}, which evolve in time according to their respective hydrodynamic equations.

Motivated by this, we argue that, for large times, {\it all contributions of the generating function to the integral in Eq.~\eqref{eq:prob_typical} for $\Gamma=0$ are captured by fluctuating variables $\varrho_i(x,t)=\hat{\varrho}_i(\tau^{1/z}x,\tau t)$, $\tau \gg 1$, that can be obtained by propagating initial typical fluctuations:
\begin{equation}
\varrho_i(x,0)\simeq\rho_i+\tau^{-1/2z}\delta\varrho_i(x,0),
\end{equation}
where $\rho_i$ is the initial average density and $\delta\varrho_i(x,0)$ is a field with $O(1)$ fluctuations.}

The assertion that such fluctuations are always transported by the Euler equations regardless of the dynamical exponent stems from the observation that the only source of typical current fluctuations in the SCCA in the single-file limit are the interactions of ballistically-propagating initial typical fluctuations. A similar observation was recently also made for Dirac fluids in Ref.~\cite{gopalakrishnan2024nongaussian}. This is in fact in the same spirit as {\it diffusion from convection}, which is known to be the only contribution to the diffusion constants in integrable systems \cite{DeNardis2018PRL,de_nardis_diffusion_2019,Gopalakrishnan2018} and is generically also present even in non-integrable systems, provided that the eigenvalues of the flux Jacobian $\mathsf{A}$ are non-degenerate \cite{Medenjak2020}. %Here, we explicitly demonstrate that this also holds in the single-file limit of the SCCA.

We illustrate the idea at half-filling $b=0$.
%and leave the case $b\neq0$ for the SM \cite{SM}.
Since in this case the dynamical exponent of charge transport is $z=2$, the fluctuating fields $\varrho_i(x,t)$ need to be rescaled to the diffusive coordinates $\varrho_i(x,t)=\hat{\varrho}_i(\sqrt{\tau}x,\tau t)$ with the associated currents $j_i(x,t)=\sqrt{\tau}\hat{\jmath}_i(\sqrt{\tau}x,\tau t)$. The fields $\varrho_i(x,t)$ then fluctuate initially as $\varrho_i(x,0)\simeq \rho_i+\tau^{-1/4}\delta\varrho_i(x,0)$, where $\delta\varrho_i(x,0)$ is Gaussian distributed as $\langle\delta\varrho_i(x,0)\delta\varrho_j(y,0)\rangle=\mathsf{C}_{ij}\delta(x-y)$ with the susceptibility matrix $\mathsf{C}_{ij}=\p\rho_i/\p\beta^j$ and $\rho_i$ is the background density.
%(see SM for the explicit expression).
The fluctuating particle fields $\delta\varrho_\pm(x,0)$ are then transported by the Euler equation $\p_t\delta\varrho_\pm(x,t)\pm\sqrt{\tau}\p_x\delta\varrho_\pm(x,t)=0$, which can be solved as $\delta\varrho_\pm(x,t)\simeq \rho_{\pm}+\tau^{-1/4}\delta\varrho_\pm(x\mp \sqrt{\tau} t,0)$. Likewise the charge field $\delta\varrho_c(x,t)$ also fluctuates as $\delta\varrho_c(x,t)\simeq \tau^{-1/4}\delta\varrho_c(x-X_d(t),0)$ where $X_d(t)$ is the characteristics that satisfies $\dd X_d(t)/\dd t=\sqrt{\tau}v(X_d(t),t)$. With these we obtain the leading behavior of the fluctuating charge current at the origin $j_c(0,t)\simeq \tau^{-1/4}\p_t\int_0^{X_d(t)}\dd x\,\delta\varrho_c(-x,0)$, which in turn allows us to write down the functional integral that describes the contribution to the typical fluctuations by these fields 
%\zk{Possible to just indicate substitutions inserted into Eq.~5?}\ty{[we refer to some of the structures in the integral later in the main text, so it looks a bit difficult to remove this]}
\begin{align}
    \langle e^{\lambda \hat{J}_c(\tau)}\rangle&\simeq\frac{1}{Z}\int_{(\mathbb{R})}\mathcal{D}\underline{\delta\varrho}(\cdot,0) e^{-\frac{\mathsf{C}^{ij}}{2}\int_\mathbb{R}\dd x\delta\varrho_i(x,0)\delta\varrho_j(x,0)}\n
    &\quad\times e^{\lambda \tau^{1/4}\int_0^{X_d(1)}\dd x\,\delta\varrho_c(-x,0)}, \label{eq:typ_fluc_integ2}
\end{align}
where $\mathsf{C}^{ij}=[\mathsf{C}^{-1}]_{ij}$ and $Z$ is the normalization factor of the initial probability distribution.
%that can be determined explicitly.
The integral is performed exactly in Methods, yielding the distribution
\begin{equation}\label{eq:b0_typ}
    \mathcal{P}_\mathrm{typ}(j)=\frac{1}{2\pi\Delta}\int_\mathbb{R} \frac{\dd u}{\sqrt{|u|}}\,e^{-\frac{u^2}{2\Delta^2}-\frac{j^2}{2|u|}}
\end{equation}
%\zk{Wrote in terms of M-Wright and put inline} $\mathcal{P}_\mathrm{typ}(j) = \sigma^{-1}M_{1/4}(2j/\sigma)$ with $\sigma^4 = 2\Delta^2$, where $M_{1/4}= \int_{\mathbb{R}}\frac{\dd y}{2\pi} |y|^{-1/2}\exp\left(-\frac{y^2}{4} - \frac{x^2}{4|y|}\right)$ is an M-Wright function \cite{Mainardi2020}, ----
% \begin{equation}
%     \mathcal{P}_\mathrm{typ}(j)=\frac{1}{2\pi\Delta}\int_\mathbb{R}\frac{\dd u}{\sqrt{|u|}}e^{-\frac{u^2}{2\Delta^2}-\frac{j^2}{2|u|}}, \label{sf_dist} 
% \end{equation}
which matches the result reported in Ref.~\cite{Krajnik2022}. 
The breakdown of Gaussianity %in Eq.~\eqref{sf_dist}
can be understood by noting which fluctuating fields are responsible for charge current fluctuations. At half-filling, it follows from Eq.~\eqref{eq:typ_fluc_integ2} that charge fluctuations $\delta\varrho_c(x,t)$ that travel along the characteristics $X_d(t)$ and cross $x=0$ between $t=0$ and $t=\tau$ contribute to current fluctuations. Since charge fluctuations are transported non-linearly, Gaussianity of inital fluctuations is no longer preserved and typical fluctuation become non-Gaussian. The situation changes dramatically away from half-filling, i.e. for $b\neq0$ where only particle fluctuations $\delta\varrho_\pm(x,t)$ emanating along light cones contribute to charge current fluctuations. Since particle fluctuations are transported linearly, Gaussianity of initial fluctuations is preserved by the dynamics, giving rise to Gaussian typical fluctuations (see Methods for details).

\section{Typical fluctuations of stochastic dynamics}
The distribution %\eqref{sf_dist}
features a cusp at the origin, see right panel of Figure~\ref{fig:typical_distributions}. Numerical simulations for $\Gamma >0$ at half-filling indicate that the cusp is rounded off with stochastic crossings while the distribution remains non-Gaussian.
This motivates a generalization %of the single-file distribution \eqref{sf_dist}
of the form
\begin{equation}
	\mathcal{P}^{[\Gamma]}_{\rm typ}(j) = \frac{1}{2\pi \sigma}\int_\mathbb{R} \frac{\dd u}{\sqrt[4]{u^2 + \omega^2}} e^{-\frac{u^2}{2\sigma^2} - \frac{j^2}{2\sqrt{u^2 + \omega^2}}}, \label{generalized_dist}
\end{equation}
which reduces to the single-file distribution %\eqref{sf_dist}
as $\omega \to 0$ and to a Gaussian distribution (with variance $\omega$) as $\omega \to \infty$. Treating $\omega$ and $\sigma$ as $\Gamma$-dependent fitting parameters, the distribution \eqref{generalized_dist} captures the asymptotic typical distribution for $\Gamma  \geq 0$ across 5 order of magnitude as shown in Fig.~~\ref{fig:typical_distributions}.
%\zk{Could gain some space by combining (11) and (12), but sequencing would be a bit forced.} \ty{[It looks difficult to combine both eq 11 and 12 in one equation]}

\section{Conclusions}
We have studied anomalous charge current fluctuations in a family of stochastic cellular automata and characterized both large and typical fluctuations based only on Euler hydrodynamics, reproducing the microscopic results obtained in Ref.~\cite{Krajnik2022} when specializing to the  deterministic limit $\Gamma=0$. The key insight is that initial fluctuations, carried by Euler hydrodynamics, fully control fluctuations both on the typical and large scales in the deterministic case. To implement the idea, we employed BMFT to compute the SCGF and proposed a BMFT-based approach to describe typical fluctuations induced by initial Gaussian fluctuations (in Ref.~\cite{diffusion_BMFT}, it is also shown how BMFT provides the full diffusive-order hydrodynamic equation, where the long-range correlations predicted by BMFT are the source of diffusive-order terms). Following Ref.~\cite{gopalakrishnan2024nongaussian}, our work further clarifies the hydrodynamic origin of anomalous fluctuations, which we believe is generic among systems that exhibit anomalous fluctuations.

While initial fluctuations remain the only source of large fluctuations even in the stochastic case $\Gamma\neq0$, it turns out that typical fluctuations are nontrivially influenced by the presence of stochasticity, which is not prescribed by initial fluctuations. We expect that stochasticity induces normal diffusion, and it would be very interesting to reproduce Eq.~\eqref{generalized_dist} by extending the formalism we introduced to incorporate normal diffusion. The observation of the single-file distribution in a classical spin chain \cite{Krajnik2024_1} therefore supports the absence of normal diffusion in the easy-axis regime \cite{Prelovsek1,Prelovsek2}. It would also be illuminating to obtain the exact FCS of the SCCA with generic $\Gamma$ via microscopic computations and to establish a scattering description of thermodynamics, as recently obtained for classical integrable spin chains \cite{bastianello2024landaulifschitz}.

Another important direction to pursue is to apply the BMFT-based framework we have formulated to anomalous typical fluctuations in the isotropic Heisenberg chain. We expect that such an approach would also allow us to describe the hydrodynamic origin of the superdiffusive
spin current fluctuations in the model, which have recently been scrutinized extensively.

%\ty{[Another important achievement of our work is to establish a hydrodynamic framework for characterizing anomalous typical fluctuations in integrable systems, including, in particular, the isotropic Heisenberg chain, which opens the doors for a systematic study of the superdiffusive partial KPZ physics observed therein.]}

\section{Acknowledgements}
We thank Bruno Bertini, Benjamin Doyon, Sarang Gopalakrishnan, Enej Ilievski, Katja Klobas, Marcin Mierzejewski, Peter Prelovšek, Tomaž Prosen, and Herbert Spohn for useful discussions. We are also grateful to Alvise Bastianello for his comments on the manuscript.
\v{Z}K is supported by the Simons Foundation as a Junior Fellow of the Simons Society of Fellows (1141511). This paper is dedicated to the memory of Marko Medenjak (1990-2022).

\appendix
\section{Solution of the BMFT equations}\label{sec:deriv_bmft}
As outlined in the main text, BMFT allows us to write the generating function $ \langle e^{\lambda \hat{J}(\tau)}\rangle$ as a functional integral
\begin{align}
    \langle e^{\lambda \hat{J}(\tau)}\rangle&=\int_{(\mathbb{S})}\mathcal{D}\underline{\varrho}(\cdot,\cdot)e^{-\tau\mathcal{F}[\underline{\varrho}(\cdot,0)]}\delta(\p_t\underline{\varrho}+\p_x\underline{j})\n
    &=\int_{(\mathbb{S})}\mathcal{D}\underline{\varrho}(\cdot,\cdot)\mathcal{D}\underline{H}(\cdot,\cdot)e^{-\tau S[\underline{\varrho},\underline{H}]},
\end{align}
where the rate function $\mathcal{F}[\underline{\varrho}(\cdot,0)]$ reads \cite{BMFT}
\begin{equation}\label{eq:bmft_initial_fluc}
    \mathcal{F}[\underline{\varrho}(\cdot,0)]=\int_\mathbb{R}\dd x\,\left((\beta^i(x,0)-\beta^i)\varrho_i(x,0)-f+f(x,0)\right),
\end{equation}
with Lagrange multipliers $\beta^i(x,0)$, the free energy density $f(x,0)$ associated with the initial fluctuating density profile $\underline{\varrho}(\cdot,0)$, and the background free energy $f$. The long-time behavior of the generating function is thus ruled entirely by the saddle point of the action $S[\underline{\varrho},\underline{H}]$ given by Eq.~\eqref{eq:SCGF_action}. Such a saddle point is characterized by the saddle point equation $\delta S[\underline{\varrho},\underline{H}]=0$, which can be recast into the following BMFT equations
\begin{align}
	&\partial_t \beta^\pm (x, t) \pm \partial_x \beta^\pm (x, t) \pm  b_\pm(x,t)\partial_x \beta^c(x, t) = 0, \label{eq1}\\
	&\partial_t H^\pm (x, t) \pm \partial_x H^\pm (x, t)  \pm  b_\pm(x, t) \partial_x \beta^c(x, t)= 0, \label{eq2}\\
	&\partial_t \beta^c (x, t) +  v(x, t)\partial_x  \beta^c (x, t) = 0, \label{eq3}\\
	&\partial_t H^c (x, t) + v(x, t) \partial_x H^c (x, t) = 0, \label{eq4}
\end{align}
with the boundary conditions at $t=0$ and $t=1$
\begin{align}
	&H^\pm(x,0) = \beta^\pm(x, 0) - \beta^\pm, \label{cond1}\\
	&H^\pm(x,1) = 0, \label{cond2}\\
	&H^c(x,0) = \beta^c(x, 0) - \beta^c + \lambda \Theta(x), \label{cond3}\\
	&H^c(x,1) = \lambda \Theta(x). \label{cond4}
\end{align}
where $\Theta(x)$ is the Heaviside step function and $b_\pm(x,t):=b(x, t) \varrho_\mp(x, t)/\varrho(x, t)$.

To solve the BMFT equations, we first start by noting that Eq.~\eqref{eq4} can be solved using the method of characteristics as $H^c(x, t) = \lambda \Theta(x- x_2(t))$, where the characteristic line $x_2(t)$ is defined as the solution to $\dd x_2(t)/\dd t=v(x_2(t),t)$ with $x_2(1)=0$. With the boundary condition this then gives another boundary condition $\beta^c(x, 0) = \beta^c + \lambda \left[\Theta(x-x_2(t)) - \Theta(x)\right]$, which can be used to obtain the full profile of $\beta^c(x, t)$
\begin{equation}
    \beta^c(x, t) = \beta^c + \lambda \left[\Theta(x-x_2(t)) - \Theta(x-x_1(t))\right],
\end{equation}
where another characteristic line $x_1(t)$ satisfies $\dd x_1(t)/\dd t=v(x_1(t),t)$ with $x_1(0)=0$.

Next we turn to the equations for right/left movers, whose characteristics $x_\pm(t)$  are trivial
\begin{equation}
	x_\pm(t) = x \pm t
\end{equation}
along which the fields $H^\pm(x, t)$ evolve according to
\begin{equation}
	\tfrac{\dd }{\dd t} H^\pm(x_\pm(x, t)), t) = \mp \lambda b_\pm(x_\pm(t), t) \delta(x_\pm(t) - x_2(t)). \label{Hpm_evolution}
\end{equation}
Integrating the boundary condition $H^\pm(x,1)=0$ backwards using Eq.~\eqref{Hpm_evolution}, we then obtain $H^\pm(x, t) = \lambda b_2 \left[\Theta(x_2(t)-x) - \Theta(\mp(\tau-t)-x) \right]$, where we noted that $b(x_i(t), t) = b_i$ is constant along $x_i(t)$. We emphasize that $b_2$ cannot be identified with $\tanh \beta^c(0, \tau)$, the value of $\beta^c$ at that point being singular due to the incidence of the contour $x_2$. However, the solution \eqref{eq2} shows that $H^\pm(x, 0)$ are constant on the intervals $(-1, x_2(0))$ and $(x_2(0), 1)$ respectively. To determine its value, we note that  the pair of Eqs.~\eqref{eq2} can be rewritten as $\tfrac{\dd }{\dd t}\left[H^\pm(x_\pm(t), t) + \log \cosh \beta^c(x_\pm(t), t) \right] = \mp \lambda b_1 \delta(t - t_1)$, which we use to propagate the initial condition \eqref{cond2} from $t=1$ back to $t=0$. With Eq.~\eqref{eq2} and the boundary condition Eq.~\eqref{cond1} we can now fully specify $\beta^\pm(x, 0)$. Invoking a similar relation $\tfrac{\dd }{\dd t}\left[\beta^\pm(x_\pm(t), t) + \log \cosh \beta^c(x_\pm(t), t) \right] = 0$, we are finally in the position to obtain the full profile of $\beta^\pm(x,t)$:
\begin{align}
    &\beta^\pm(x,t)=\beta^\pm+s[\Theta(\pm t-x)-\Theta(\pm(t-1)-x) \n
    &\quad-\Theta(x_1(t)-x)+\Theta(x_2(t)-x)]\frac{\cosh \beta^c(\lambda)}{\cosh \beta^c},
\end{align}
where  $\beta^c(\lambda) := \beta^c + \lambda s$ and $s :=  {\rm sgn}(x_1-x_2)$. It is a simple matter to argue that the sign $s$ remains constant in time, and is given by $s=\mathrm{sgn}(\lambda b)$. For $\beta^\pm = \beta$ and $t\in(0,1)$ the solution is time-independent along the ray $x=0$ and reads
\begin{align}
\beta^{(\lambda),c}(0, t) &= \beta^c(\lambda) = \beta^c + |\lambda|{\rm sgn}\,b \label{eq:bmft_sol_charge},\\
   \beta^{(\lambda),\pm}(0, t) &= \beta - 2\Theta(\mp \lambda b)\log \left[\frac{\cosh \beta^c(\lambda)}{\cosh \beta^c} \right]. \label{eq:bmft_particle_charge}
\end{align}
In terms of the solution $\underline{\beta}^{(\lambda)}(x,t)$ of the BMFT equations, the SCGF is computed as \cite{BMFT}
\begin{equation}\label{eq:SCGF_BMFT}
    F(\lambda)=\int_0^\lambda\dd \lambda'\int_0^1\dd t\,j_c[\underline{\varrho}^{(\lambda')}(0,t)]
\end{equation}
where $\underline{\varrho}^{(\lambda)}(x,t)=\underline{\varrho}[\underline{\beta}^{(\lambda)}(x,t)]$. Plugging Eqs.~\eqref{eq:bmft_sol_charge} and \eqref{eq:bmft_particle_charge} into Eq.~\eqref{eq:SCGF_BMFT}, we obtain Eq.~\eqref{eq:SCGF_exact}.
To our knowledge, the above is the first exact solution of full BMFT equations in an interacting model and therefore serves as an independent check of their validity.

\section{Comparison between BMFT and BFT}  
It was argued in Ref.~\cite{BMFT} that BMFT in fact reduces to ballistic fluctuation theory (BFT) \cite{Doyon2019,Myers2020,DVDV2023} when the former is specialized to equilibrium. Let us demonstrate that this is indeed the case in the SCCA as well. BFT is a large deviation theory that describes the current fluctuations of the system that supports ballistic transport. A major difference between BFT and BMFT is that BFT is applicable only to systems in equilibrium, which makes the structure of BFT easier to handle. The main observation in BFT is that the insertion of the operator $e^{\lambda \hat{J}(t)}$ in a GGE amounts to a modification of the GGE parameterized by the Lagrange multipliers $\underline{\beta}$ into that parameterized by the $\lambda$-dependent Lagrange multipliers $\underline{\beta}(\lambda)$. Adopting to the charge transport in the SCCA, such multipliers $\underline{\beta}(\lambda)$ satisfy the so-called flow equation
\begin{equation}\label{flow_beta}
    \p_\lambda\beta^i(\lambda)=[\mathrm{sgn}\mathsf{A}(\lambda)]_{c}^{~i},
\end{equation}
where $\mathsf{A}(\lambda)$ is the flux Jacobian evaluated with respect to the multipliers $\underline{\beta}(\lambda)$. More explicitly, the equations for multipliers read
\begin{align}
\p_\lambda\beta^\pm(\lambda)&=\pm2b\Theta(\mp(\beta^+(\lambda)-\beta^-(\lambda))), \label{flow_beta2}\\
\p_\lambda\beta^c(\lambda)&=\mathrm{sgn}(\beta^+(\lambda)-\beta^-(\lambda)). \label{flow_beta3}
\end{align}
In terms of the solution of the flow equation, the SCGF is given by
\begin{equation}\label{bft_scgf}
F(\lambda)=\int_0^\lambda\dd\lambda'\,j_c(\lambda'),
\end{equation}
where $j_c(\lambda)$ is the charge current evaluated again with respect to $\underline{\beta}(\lambda)$. The flow equations Eqs.~\eqref{flow_beta2} and \eqref{flow_beta3} can be in fact solved explicitly (for $\beta^\pm = \beta)$, and the solution turns out to precisely coincide with the solution of the BMFT equations at $x=0$ Eqs.~\eqref{eq:bmft_sol_charge} and \eqref{eq:bmft_particle_charge}. We thus confirm that the modified $\lambda$-dependent GGE is indeed realized along the ray $x=0$ as proposed in Ref.~\cite{BMFT}.

\section{Typical probability distribution at $\Gamma=0$}
In order to evaluate the functional integral Eq.~\eqref{eq:typ_fluc_integ2}, we first need to recast it in terms of normal modes $\delta\varrho=R^{-1}\delta n$ such that $R\mathsf{C}R^{-1}=I$, where the transformation matrix $R^{-1}$ reads
\begin{equation}
    R^{-1}=\frac{1}{\sqrt{2}}\begin{pmatrix}
       \Delta_+ & 0 & 0\\
        0 &   \Delta_- & 0\\
        b \Delta_+ & b \Delta_- & \sqrt{(1-b^2)(\rho_++\rho_-)}
    \end{pmatrix},
\end{equation}
with $\Delta_\pm= \sqrt{\rho_\pm(1-\rho_\pm)}$. Using this, we can evaluate the normalization factor
\begin{align}
Z&=\int_{(\mathbb{R})}\mathcal{D}\underline{\delta\varrho}(\cdot,0) e^{-\frac{\mathsf{C}^{ij}}{2}\int_\mathbb{R}\dd x\delta\varrho_i(x,0)\delta\varrho_j(x,0)}\n
&=(2\pi)^{3/2}\det R^{-1}=\sqrt{2\pi(1-b^2)\rho}\Delta^2,
\end{align}
which allows us to rewrite Eq.~\eqref{eq:typ_fluc_integ2} as
\begin{align}
     \langle e^{\lambda \hat{J}_c(\tau)}\rangle&\simeq
     \frac{1}{(2\pi)^{3/2}}\int_{(\mathbb{R})}\mathcal{D}\delta\underline{n}(\cdot,0)  e^{-\frac{1}{2}\int_\mathbb{R}\dd x(\delta n_i(x,0))^2} \n
     &\quad\times e^{\lambda \tau^{1/4}\sqrt{\rho}\int\dd x\chi(-|X_d(1)|<x<0)\delta n_c(x,0)]}\n
     &=\frac{1}{2\pi}\int_{(\mathbb{R})}\mathcal{D}\delta n_1(\cdot,0)\mathcal{D}\delta n_2(\cdot,0)  \n
     &\quad\times e^{-\frac{1}{2}\int\dd x[(\delta n_1(x,0))^2+(\delta n_2(x,0))^2]}e^{\frac{1}{2}\lambda^2\tau^{1/2}\rho|X_d(1)|}
\end{align}
Assuming $X_d(t)$ is monotonic in time, we then have
\begin{align}
     &\langle e^{\lambda \hat{J}_c(\tau)}\rangle\n
     &\simeq\int_{(\mathbb{R})}\frac{\mathcal{D}\delta n_1(\cdot,0)\mathcal{D}\delta n_2(\cdot,0)}{2\pi} e^{-\frac{1}{2}\int_\mathbb{R}\dd x[(\delta n_1(-x,0))^2+(\delta n_2(x,0))^2]}\n
     &\quad\times e^{\frac{\Delta\lambda^2 \tau^{1/4}}{2\sqrt{2}}|\int_0^{\sqrt{\tau}}\dd x\,\delta n_1(-x,0)-\delta n_2(x,0)|}.
\end{align}
Introducing the variables $\delta n(x,0)=(\delta n_1(-x,0)+\delta n_2(x,0)/\sqrt{2}$ and $\delta\Bar{n}(x,0)=(\delta n_1(-x,0)-\delta n_2(x,0))/\sqrt{2}$, we can carry out the integral explicitly
\begin{align}
    \langle e^{\lambda \hat{J}_c(\tau)}\rangle 
&\simeq\frac{1}{\sqrt{2\pi}}\int_{(\mathbb{R})}\mathcal{D}\delta \Bar{n}(\cdot,0) e^{-\frac{1}{2}\int_\mathbb{R}\dd x(\delta \Bar{n}(x,0))^2}\n 
&\quad\times e^{\frac{\Delta\lambda^2 \tau^{1/4}}{2}|\int_0^{\sqrt{\tau}}\dd x\,\delta \Bar{n}(x,0)|}\n
     &=e^{\xi^4/2}(1+\mathrm{erf}(\xi^2/\sqrt{2})),
\end{align}
where $\xi^2=\Delta\lambda^2\sqrt{\tau}/2$ and we invoked the identity $\int_\mathbb{R}\dd y\,e^{-x|y|-(y/2)^2}=2\sqrt{\pi}e^{x^2}\mathrm{erfc}\,x$. This in turn gives the non-Gaussian typical distribution Eq.~\eqref{eq:b0_typ}. %Eq.~\eqref{sf_dist}.

The case of $b\neq0$ can be dealt with in a similar way.
Charge transport now has the dynamical exponent $z=1$, thus the initial typical fluctuations are prescribed by the fluctuating fields $\varrho_i(x,0)=\hat{\varrho}_i(\tau x,0)$
\begin{equation}
    \varrho_i(x,0)\simeq \rho_{i}+\tau^{-1/2}\delta\varrho_i(x,0),
\end{equation}
where $\delta\varrho_i(x,0)$ is normal distributed with the susceptibility $\mathsf{C}_{ij}$ as before.
Since such fluctuations are disseminated by the Euler equation, the left and right movers remain diffusively fluctuate
\begin{equation}\label{eq:fluc_pm}
\varrho_\pm(x,t)\simeq \rho+\tau^{-1/2}\delta\varrho_\pm(x\mp t,0).
\end{equation}
We thus realize that the Euler-scale charge current $j_c(0,t)$ has the leading behavior
\begin{align}
     j_c(0,t)&=b(0,t)\frac{\varrho_+(0,t)-\varrho_-(0,t)}{2}\n
&\simeq b\tau^{-1/2}\frac{\delta\varrho_+(-t,0)-\delta\varrho_-(t,0)}{2},
\end{align}
where in the second line we used Eq.~\eqref{eq:fluc_pm}.
Note that the fluctuation of the normal mode $b(x,t)$ does not contribute here.
With these, we can write down the cumulant generating function as a Gaussian integral
\begin{align}
     \langle e^{\lambda \hat{J}_c(\tau)}\rangle&\simeq\frac{1}{Z}\int_{(\mathbb{R})}\mathcal{D}\underline{\delta\varrho}(\cdot,0) e^{-\frac{\mathsf{C}^{ij}}{2}\int_\mathbb{R}\dd x\delta\varrho_i(x,0)\delta\varrho_j(x,0)}\n
    &\quad\times e^{\frac{b\lambda\sqrt{\tau}}{2}\int_0^1\dd t\,(\delta\varrho_+(-t,0)-\delta\varrho_-(t,0))} \n
    &=\frac{1}{2\pi}\int_{(\mathbb{R})}\mathcal{D}\underline{\delta n}(\cdot,0)\,e^{-\int_\mathbb{R}\dd x\frac{1}{2}(\delta n_i(x,0))^2} \n
    &\quad\times e^{-\int_\mathbb{R}(\mu_1(x)\delta n_1(x,0)+\mu_2(x)\delta n_2(x,0))}
    \label{eq:typ_fluc_integ1_SM}
\end{align}
where
\begin{align}
     \mu_1(x)=\mu\chi(-1<x<0),\quad 
     \mu_2(x)=-\mu\chi(0<x<1)
\end{align}
with $\chi(x)$ the indicator function and $\mu=\lambda b\sqrt{\tau}\Delta/2$. Performing the integral explicitly gives $\langle e^{\lambda \hat{J}(\tau)}\rangle\simeq e^{\mu^2/2}$.
Transforming back to the probability distribution  $\mathcal{P}(j\tau^{-1/2z}|\tau)=(2\pi)^{-1}\int_\mathbb{R}\dd\lambda\,e^{-\ii\lambda j\tau^{-1/2z}}\langle e^{\ii\lambda J(\tau)}\rangle$, we finally get
\begin{equation}
   \mathcal{P}^{[0]}_\mathrm{typ}(j)=\frac{1}{\sqrt{2\pi\sigma^2}}e^{-\frac{j^2}{2\sigma^2}},\quad \sigma^2=(b\Delta)^2.
\end{equation}

\begin{figure}[t]
    \subfloat[\label{fig:bneq0}]{%
        \includegraphics[width=0.5\columnwidth]{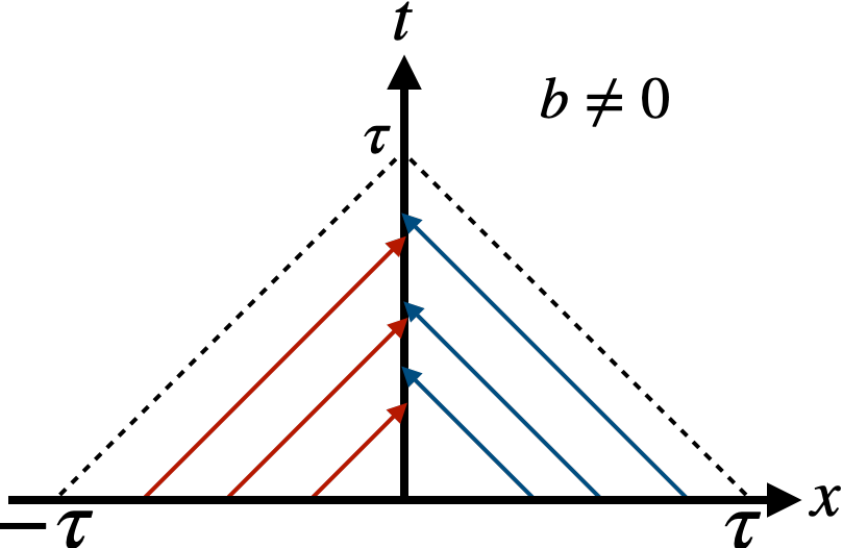}%
    }\hspace*{\fill}
    \subfloat[\label{fig:b0}]{%
        \includegraphics[width=0.5\columnwidth]{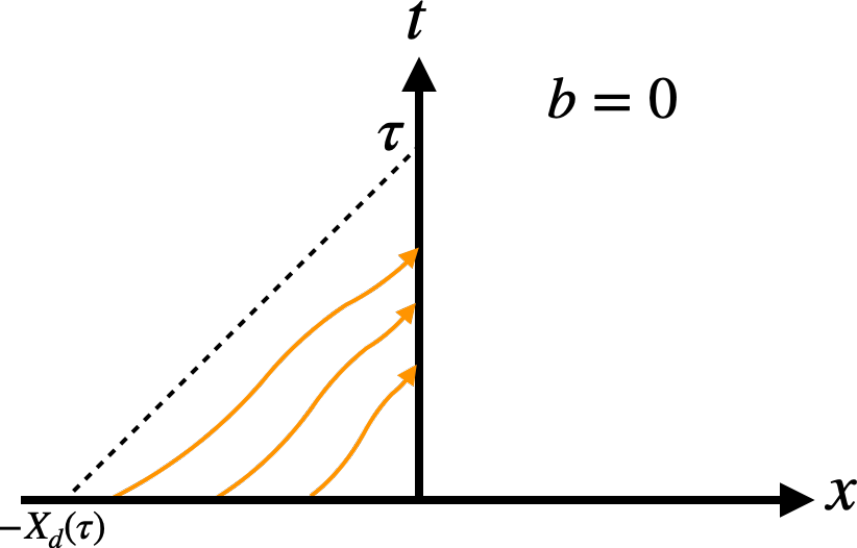}
    }
    \caption{Initial fluctuations that influence the charge current fluctuations. (a): when $b\neq0$, the particle fluctuations $\delta\varrho_\pm(x,0)$ that cross $x=0$ between $t=0$ and $t=\tau$ give rise to the current fluctuations. (b): when $b=0$ the charge fluctuations $\delta\varrho_c(x,0)$ that cross $x=0$ between $t=0$ and $t=\tau$ contribute to the current fluctuations. Note that while the trajectory is now nonlinear, the slope cannot exceed $1$ because $|\dd X_d(t)/\dd t|<1$. Here we assumed $X_d(t)>0$ for $t\in[0,\tau]$.}
\end{figure}

As we discussed in the main text, we can make sense of the markedly different behaviors in the two cases $b\neq0$ and $b=0$ by appreciating which fluctuating fields are responsible for the charge current fluctuations. First, when $b\neq0$, it can be readily seen in Eq.~\eqref{eq:typ_fluc_integ1_SM} that the charge current fluctuations are completely determined by the initial particle fluctuations $\delta\varrho_\pm$. Since particle fluctuations are propagated freely in time (see Fig.~\ref{fig:bneq0}), the initial Gaussian fluctuations remain intact and as a result the typical fluctuations keep Gaussianity. This is in stark contrast to what happens at half-filling where the current fluctuations are now given solely by the charge fluctuations that are transported nonlinearly with the velocity $v$ (see Fig.~\ref{fig:b0}) according to Eq.~\eqref{eq:b0_typ}.
Gaussianity is therefore not preserved during the dynamics, resulting in the non-Gaussian distribution.

\bibliography{bib.bib}

\end{document}